\documentclass{article}

\usepackage{amsmath}
\usepackage{authblk}
\usepackage[]{graphicx}
\usepackage{color}
\usepackage[latin1]{inputenc}
\usepackage{marginnote}
\usepackage[UKenglish]{babel}

\usepackage{mathbbol}
\usepackage{amssymb}             % AMS Math
\usepackage{textcomp}

\begin{document}

\title{X-ray phase contrast topography to measure the surface stress and bulk strain in a silicon crystal}
\author{E.Massa}
\author{C.P.Sasso}
\author{M.Fretto}
\author{L.Martino}
\author{G.Mana}
\affil{INRIM -- Istituto Nazionale di Ricerca Metrologica, strada delle cacce 91, 10135 Torino, Italy. email: \textit{c.sasso@inrim.it}}

%\keyword{\LaTeX}
%\keyword{class file}
%\keyword{documentation}
\maketitle

\begin{abstract}
The measurement of the Si lattice parameter by x-ray interferometry assumes the use of strain-free crystals, which might not be true because of intrinsic stresses due to surface relaxation, reconstruction, and oxidation. We used x-ray phase-contrast topography to investigate the strain sensitivity to the finishing, annealing, and coating of the interferometer crystals. We assessed the topography capabilities by measuring the lattice strain due to films of copper deposited on the interferometer mirror-crystal. A byproduct has been the measurement of the surface stresses after complete relaxation of the coatings.
\end{abstract}

\section{Introduction}

We used a single-crystal x-ray interferometer to measure the lattice parameter of silicon to within a fractional uncertainty approaching 1 nm/m \cite{Massa_2015}. The interferometer was ground by diamond tools and chemically etched to remove the surface damage \cite{Zawisky_2010}. If too little material is etched away, lattice strains prevent the interferometer operation; if too much, the interferometer geometry degrades, and the fringe contrast is lost.

Phase-contrast topography by x-ray interferometry is a well known tool to study defects and strains in single crystals \cite{Bonse:1976,Ohler:1999,Fodchuk_2003,Pushin:2007,Miao:2016}. We used it to investigate the effect of the surface finishing on the interferometer operation \cite{Bergamin_2000}. Our goal was to optimize the manufacturing and to trade-off between no surface damage (via chemical etching) and accurate geometry (via mechanical grinding). The test interferometers were etched step-by-step and etching was stopped when it neither improved the fringe contrast nor reduce the lattice strain. The procedure that was found optimal prescribes a first chemical etching, then machining with the finest grit size to correct about 10 $\mu$m etch errors, and a final etching for a depth of about 50 $\mu$m.

These investigations did not give certainties that surface stresses due to oxidation, relaxation, and reconstruction did not affect the lattice-parameter value. The magnitude of this error was estimated by a finite element analysis, where the surface stress (a fundamental property of the crystal interface with the environment) was modelled by an elastic membrane having a hypothetical 1 N/m tensile strength \cite{Quagliotti_2013}. We also calculated the surface stress by the density functional theory \cite{Melis_2016} and found a value exceeding 1 N/m, which value potentially jeopardizes the measurement accuracy.

Prompted by this treat and the observations of increased visibility and reduced strains after annealing reported in \cite{Heacock_2018,Heacock_2019}, we carried out new topographic investigations of the annealing and etching effects of the interferometer crystals. To test the capabilities of phase-contrast topography, we measured the strain in the crystal bulk caused by nanometric films of copper deposited on one of the interferometer crystals. As a byproduct, we obtained the in-plane mean-stress in copper films on a silicon substrate.

Phase-contrast imaging proved to be an extremely sensitive technique to measure stress in films, only a few nanometres thick. Since it affects the design, processing, performance, reliability, and lifetime of advanced materials and components, the measurement of stress in thin films and coatings is also a crucial issue in materials science and technology \cite{Sharma:2015,Abadias:2018}.
\begin{figure}\center
\includegraphics[width=\columnwidth]{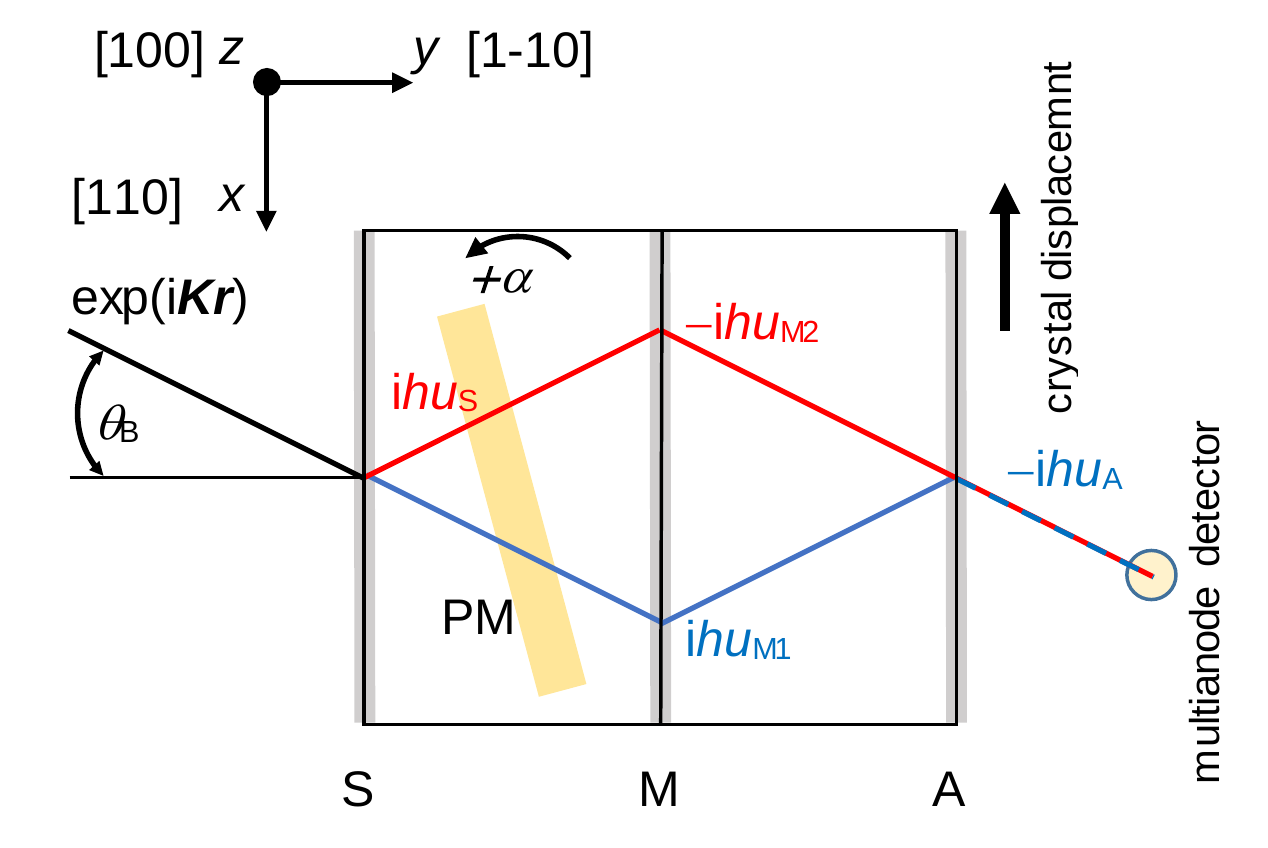}
\caption{X-ray phase-contrast topography. S, splitter; M mirror; A, analyser; PM, phase modulator. The x-ray paths are drawn in red (RRT path) and blue (TRR path). The Bragg angle is out of scale. The phase delay of each reflection is given. The x-rays crossings with the mirror are spaced by 4 mm.}
\label{fig-01}
\end{figure}
\section{Experimental set-up}

Figure \ref{fig-01} shows the apparatus for phase-contrast topography. A first crystal (splitter) splits 17 keV x-rays from a fixed anode $(0.1 \times 10)$ mm$^2$ Mo K$\alpha$ source, which are recombined, via a mirror-like crystal, by the third (analyser). X-rays are roughly collimated by a $(0.5 \times 16)$ mm$^2$ slit placed in front of the interferometer. The interference fringes are imaged onto a multianode photomultiplier tube through a vertical 15 mm pile of eight 1 mm NaI(Tl) scintillators, spaced by 1 mm shades.

\begin{figure}\center
\includegraphics[width=0.9\columnwidth]{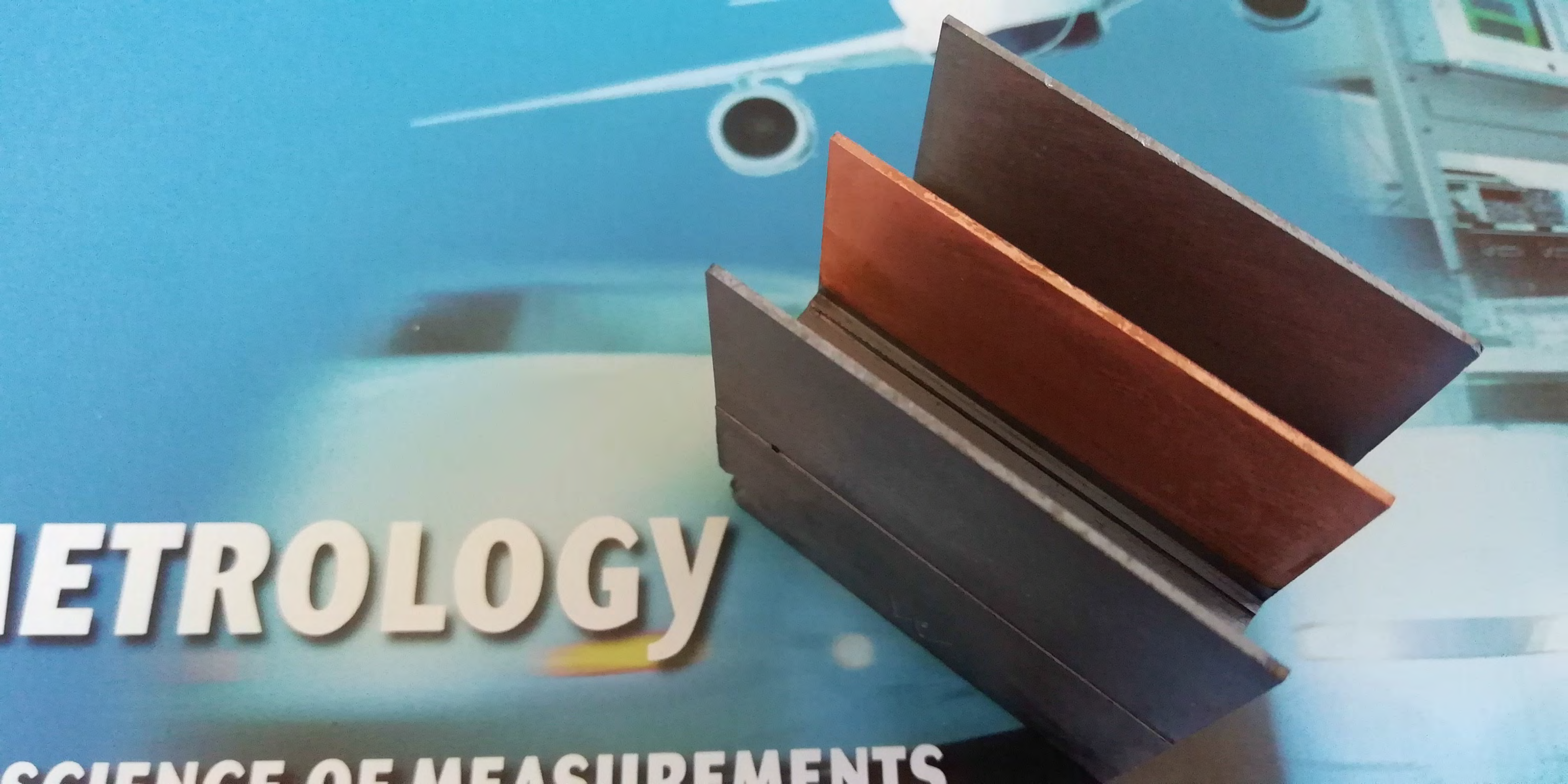}
\caption{Photograph of the x-ray interferometer with a Cu film on the mirror crystal. The film thickness has been increased to make it visible.}
\label{fig-XINT}
\end{figure}

The interferometer blades (splitter, mirror, and analyser) are $(35\times 18\times 0.8)$ mm$^3$, spaced 10 mm apart, and protrude from a common base (see Fig.\ \ref{fig-XINT}). Since the x-ray source and detector are 0.8 m and 0.3 m apart from the mirror and the beams' width at the mirror is 1 mm, the images of the scintillator pixels projected on the mirror are, on the average, $(1 \times 3)$ mm$^2$. The projected image of the 15 mm scintillator pile is 13 mm height, from the mirror top downward.

As shown in Fig.\ \ref{fig-01}, we surveyed the moir\'e pattern by shifting the interferometer in 0.5 mm steps along the $x$-axis and detecting the interference fringes in 61 adjacent $(1\times 13)$ mm$^2$ vertical (overlapping) slabs subdivided into 8 (overlapping) pixels $(1 \times 3)$ mm$^2$. Therefore, the $(61 \times 8)$ pixels survey a $(30 \times 10)$ mm$^2$ area, by using the coordinates of the pixel centres.
\section{Measurement equation}

In a geometric-optics model of the interferometer operation (with the  positive-exponent choice representing a plane wave with positive wave number $K$, see Fig.\ \ref{fig-01}), each reflection delays the x-ray phase by $\pm hu_i(x,z)$ \cite{Vittone:1997a,Vittone:1997b}, where $h=2\pi/d$ is the reciprocal vector, $d$ is the diffracting-plane spacing, $u_i(x,z)$ $(i=A, M1, M2, S)$ is the $x$ component of the displacement field of the splitter-, mirror-, or analyzer-lattice. The sign is positive if the displacement $u_i(x,z)$ occurs in the same direction as the $x$-component of the incident-beam wave vector and negative otherwise. No phase delay occurs in the transmissions.

The phase delays along the two paths reaching the observation plane -- one performing two reflections (R) followed by one transmission (T), the other one transmission followed by two reflections -- are thus
\begin{subequations}
\begin{eqnarray}
 \phi_{\rm RRT} &= &h(u_S-u_{M2}) \\
 \phi_{\rm TRR} &= &h(u_{M1}-u_A) .
\end{eqnarray}
\end{subequations}
An interference occurs because the rays overlap after crossing crystal lattices whose planes are misplaced one to the other. The total phase is
\begin{equation}\label{phi}
 \phi_u = \phi_{\rm RRT}-\phi_{\rm TRR} = h(u_S+u_A-u_{M1}-u_{M2}) .
\end{equation}
A phase modulator, plastic sheet 1 mm thick, is placed between the splitter and mirror. A simple geometrical analysis shows that -- with the positive-exponent choice to represent a plane wave with positive $K$ (see Fig.\ \ref{fig-01}) -- it varies the interference phase by
\begin{equation}\nonumber
 K(T_{\rm RRT} - T_{\rm TRR})\alpha \approx -2KT(n-1)\theta_B\alpha,
\end{equation}
where $\alpha $ is the angle of rotation, $T$ the thickness of the modulator, $T_{\rm RRT, TRR} $ the length of the X-ray path through the modulator, $n <1 $ the index of refraction, $\theta_B$ the Bragg angle and the linearization with respect to the rotation angle is valid if $\alpha \ll 1$ rad.

When the phase modulator is rotated, the moving fringes are detected by each of the eight photomultiplier channels, making it possible to extract the effective displacement field $u_x = u_S + u_A-u_{M1} -u_{M2}$. The measurement equation is
\begin{equation}\label{fringe}
 I_n = I_{0n}\left[ 1 + \Gamma_n\cos(\phi_n +\Omega\alpha) \right] ,
\end{equation}
where $n=1, 2, ... (61\times 8)$ label the image pixel, $I_{0n}$ is the average count rate, $\Gamma_n >0$ the contrast, and $\Omega = 2KT(1-n)\theta_B >0$ the period.

The phases $\phi_n \in [0,2\pi[$ in the $(61 \times 8)$ image pixels were recovered by (non-linear) least-squares estimations, with the $\Gamma_n > 0$ and $\Omega > 0$ constraints. After unwrapping, we found the optimal polynomial regression $\phi(x,z)$ explaining the $\phi_n$ data and used it to infer the effective lattice displacement $u_x(x,z)=\phi(x,z)/(2\pi)$. The trade-off between underfitting and overfitting was carried out according to \cite{Mana:2014,Mana:2019}. Eventually, we calculated  the components $\epsilon_{xx}(x,z) = \partial_x u_x(x,z)$ (normal strain, the relative variation of lattice spacing), and $\epsilon_{xz}(x,z) = \partial_z u_x(x,z)$ (shear strain, the lattice-plane rotations about the $y$-axis) of the strain tensor.

Since the fringes phase is recovered only modulo $2\pi$, a constant $u_x(x,z)$ field is undetectable. However, positive phase gradients correspond to displacements of the splitter and analyser lattices in the $x$-direction of the incident x-rays. The opposite is true for the mirror lattice. Therefore, tensile and compressive strains can be distinguished.

In (\ref{phi}), we neglected minor contributions to the phase, coming from deviations from ideally plane and parallel surfaces of the crystals and phase modulator. They are discussed in \cite{Vittone:1997b,Bergamin_2000} and might amount to a few per cent of a period. However, since we are interested in the strain changes after reprocessing of the crystal surfaces, we are looking at the difference of subsequent phase surveys and these constant contributions are mostly irrelevant.
\begin{figure}\centering
\includegraphics[width=0.9\columnwidth]{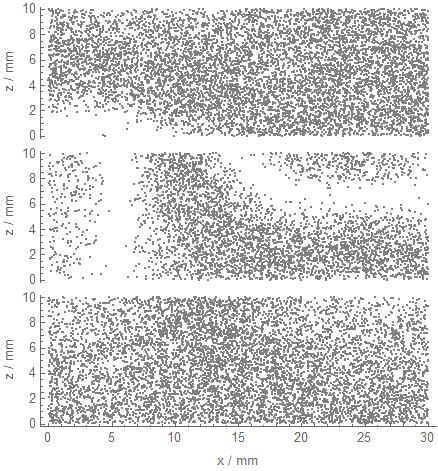}
\caption{Moir\'e topographies of the x-ray interferometer after optimal grinding and etching (top), annealing (middle), and re-etching (bottom). The coordinates are relative to the bottom left corner of the image. The fringe contrast has been artificially enhanced to one to improve visibility.} \label{fig-02}
\end{figure}
\begin{figure}\centering
\includegraphics[width=0.9\columnwidth]{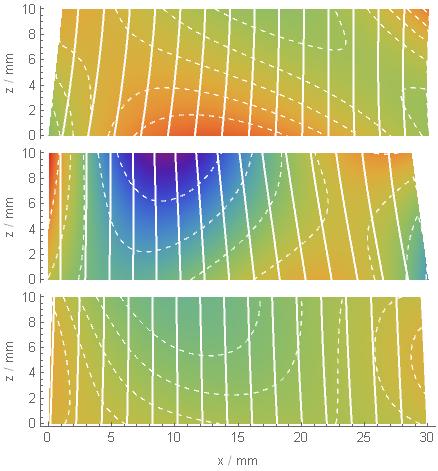}
\caption{Density plots of the $xx$ component of the strain tensor inferred from the moir\'e topographies shown in Fig.\ \ref{fig-02}. From top to bottom: optimal grinding and etching, annealing, and re-etching. The colour scale is from $-11$ nm/m (blue) to $+6$ nm/m (red). Contour lines are dashed. Solid lines are the lattice planes with distortions magnified. The coordinates are relative to the bottom left corner of the image.}\label{fig-03a}
\end{figure}
\begin{figure}\centering
\includegraphics[width=0.9\columnwidth]{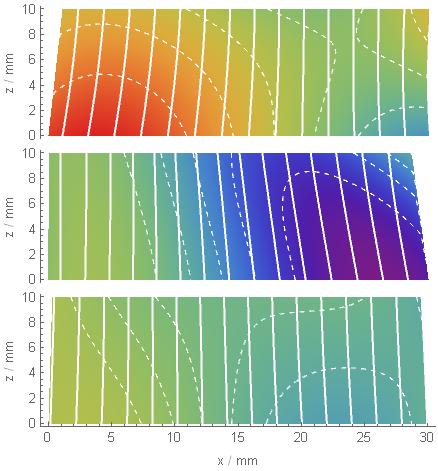}
\caption{Density plots of the $xz$ component of the strain tensor inferred from the moir\'e topographies shown in Fig.\ \ref{fig-02}. From top to bottom: optimal grinding and etching, annealing, and re-etching. The colour scale is from $-12$ nrad (blue) to $+11$ nrad (red). Contour lines are dashed. Solid lines are the lattice planes with distortions magnified. The coordinates are relative to the bottom left corner of the image.}\label{fig-03b}
\end{figure}
\begin{figure}\centering
\includegraphics[width=0.9\columnwidth]{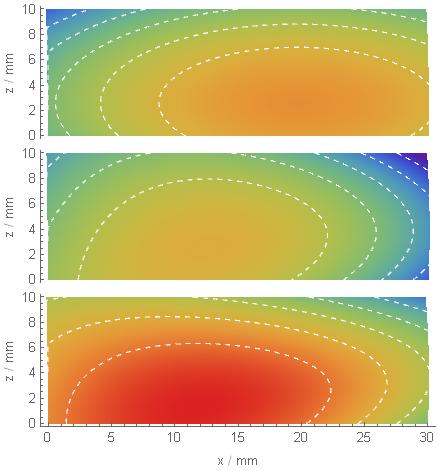}
\caption{Density plots of the fringe contrast inferred from the moir\'e topographies shown in Fig.\ \ref{fig-02}. From top to bottom: optimal grinding and etching, annealing, and re-etching. The colour scale is from $14$\% (blue) to $81$\% (red). Contour lines are dashed. The coordinates are relative to the bottom left corner of the image.}\label{fig-03c}
\end{figure}
\section{Crystal annealing}

We surveyed the lattice strain of the interferometer crystals after optimal grinding and etching. Next, the interferometer was annealed in an evacuated tube furnace (the residual pressure was $10^{-5}$ mbar), at 800 $^\circ$C for 12 h. Eventually, the finishing of the crystal surfaces was reset by re-etching. The sequence of moir\'e topographies is shown in Fig.\ \ref{fig-02}.

After finding the optimal approximations of the lattice-displacement fields (the residual standard deviations are about 3\% of the peak-to-peak displacements), we calculated the $xx$ and $xz$ components of the strain tensors. Figures \ref{fig-03a} and \ref{fig-03b} highlight that the surface finishing plays a role in determining the bulk spacing and tilt of the diffracting planes. Sorted as in Figs.\ from \ref{fig-02} to \ref{fig-03b}, the mean strains and peak-to-peak variations are
\begin{equation}\begin{array}{llll}
                  \overline{\epsilon_{xx}} =\;\;\,0_{-2}^{+4} &{\rm nm/m,}  &\overline{\epsilon_{xz}} =+4_{-5}^{+11}     &{\rm nrad} \\
                  \overline{\epsilon_{xx}} =-3_{-11}^{+6}     &{\rm nm/m,}  &\overline{\epsilon_{xz}} =-6_{-12}^{+1}     &{\rm nrad}\\
                  \overline{\epsilon_{xx}} =-1_{-4}^{+2}      &{\rm nm/m,}  &\overline{\epsilon_{xz}} =-1_{-4}^{+2}      &{\rm nrad}\
                \end{array} .
\end{equation}

While the normal strain did not change significantly, there was an overall alignment of the diffracting planes, indicating that some stress in the base was relieved. Figure \ref{fig-03c} shows that the realignment of the interferometer crystals was significant enough to improve the fringe contrast.

Since, according to (\ref{phi}), four displacement fields superimpose, we cannot give a measure of the strain in any single crystal. If we assume the four fields uncorrelated, by dividing the observed peak-to-peak strains by two, we can estimate a (local) surface effects on the spacing and tilt of the diffracting planes of about $\pm 1.5$ nm/m and $\pm 1.5$ nrad.

As regards the mean strain in a single crystal and measurement of the lattice parameter \cite{Massa_2015,Bartl_2017,Fujii_2017}, we might expect a fraction of these effects. We cannot, however, make any specific assertion about this point.

\section{Crystal coating}

To check the capabilities of phase-contrast topography and to gain some preliminary clues on the possible effects of the SiO$_2$ surface layer on the lattice parameter, we measured the crystal strain after auto-catalytical coatings of the mirror surface with nanometric films of copper.

The coating (see Fig.\ \ref{fig-XINT}) was carried out by an electroless galvanic displacement mechanism in a water solution of copper (II) nitrate, ${\rm Cu}({\rm NO}_3)_2$ ($\rho_{{\rm Cu}({\rm NO}_3)_2} = 60$ g L$^{-1}$), and ammonium fluoride, ${\rm NH}_4{\rm F}$ ($\rho_{{\rm NH}_4{\rm F}} = 30$ g L$^{-1}$). In this process, the copper plates the silicon surface and, simultaneously, the oxidised silicon is removed by ${\rm HF}^-$ to form water-soluble silicates and a clean interface between the Cu layer and the silicon crystal surface. Therefore, two processes go along: etching of the silicon surface and plating by copper. The overall stoichiometric reaction is \cite{mendel:1969}
\begin{eqnarray}\nonumber
 {\rm Si} + 2{\rm Cu}({\rm NO}_3)_2 + 6{\rm NH}_4{\rm F} \rightarrow
 \\ \label{plating}
 ({\rm NH}_4)_2 {\rm SiF}_6 + 4{\rm NH}_3\uparrow + 2{\rm Cu} + 4{\rm HNO}_3 .
\end{eqnarray}

The growth rate and quality of the Cu film depend on the solution composition and temperature. Therefore, we standardised them as well as the plating duration, 10 s, 20 s, and 40 s.

\subsection{Film-thickness measurement}

We measured the thickness of the Cu film by coating an optical polished single-crystal Si wafer whose surface was masked to coat (with times ranging from 15 s to 90 s with 15 s increments) six $(5 \times 5)$ mm$^2$ Cu pads, each bounded by a reference Cu-free area.

\begin{figure}\centering
\includegraphics[width=0.9\columnwidth]{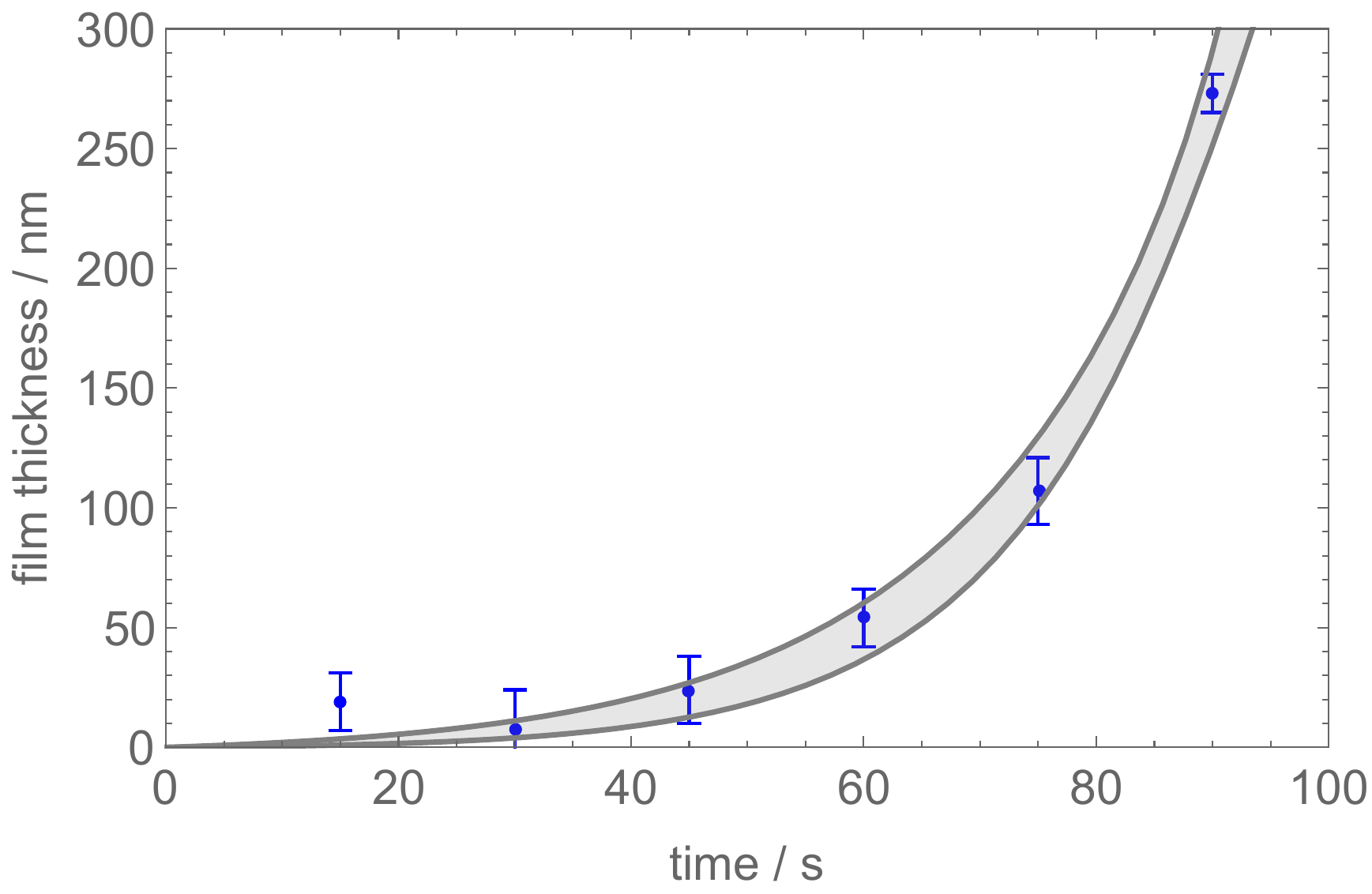}
\caption{Measured thickness of the Cu film {\it vs.} the deposition time. Bars are the 95\% confidence intervals of the data. The filled area represents the 95\% confidence intervals of the exponential fitting the data.}\label{fig-31}
\end{figure}

After the Cu films were removed by a water solution of iron(III) chloride, FeCl$_3$ ($\rho_{{\rm FeCl}_3} = 300$ g L$^{-1}$), we used an optical confocal profilometer (Sensofar S Neox Optical Profiler) to measure the steps between the coated and non-coated areas. Since the Cu molar volume is half that of Si and -- according to (\ref{plating}) -- two Cu atoms are a substitute for each Si atom removed, the thickness of the Cu films were estimated as equal to the observed drops. Figure \ref{fig-31} shows the results.

We prepared the mirror surface by the same etching/coating procedure used for the Si wafer: three increasing deposition time (10 s, 20 s, 40 s) were used to plate the surface with an increasingly thicker copper film. According to Fig.\ref{fig-31}, we estimated the thicknesses of the Cu-films on the interferometer mirror as equal to $t_{\rm Cu}=1.3(4)$ nm, $t_{\rm Cu}=4(1)$ nm, and $t_{\rm Cu}=14(3)$ nm, respectively. The parentheses are a concise notation for the standard uncertainty; the enclosed digit applies to the numeral left of themselves.
\subsection{Displacement-field measurement.}
The displacements of the mirror lattice due to the Cu coatings, $u_{\rm M}=(u_{\rm M1}+u_{\rm M2})/2$, were obtained by subtracting the regression of the pre-coating displacement field from those observed after the coatings, and reversing the sign.

\begin{figure}\centering
\includegraphics[width=0.9\columnwidth]{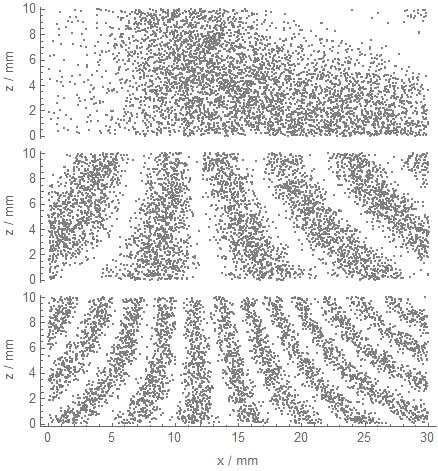}
\caption{Moir\'e topographies of the x-ray interferometer after auto-catalytical Cu plating of the mirror crystal. From top to bottom: 1.3(4) nm, 4(1) nm, and 14(3) nm film thickness. The coordinates are relative to the bottom left corner of the image. The fringe contrast has been set to one to improve visibility.}\label{fig-04}
\end{figure}

\begin{figure}\centering
\includegraphics[width=0.9\columnwidth]{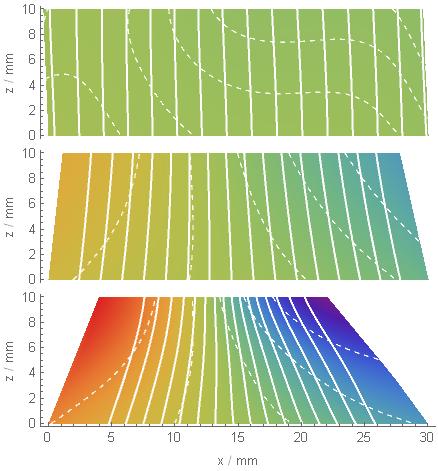}
\caption{Density plots of the $u_{\rm M}(x,z)$ values (polynomial regressions) inferred from the moir\'e topographies shown in Fig.\ \ref{fig-04}. The coordinates are relative to the bottom left corner of the image. From top to bottom: 1.3(4) nm, 4(1) nm, and 14(3) nm thickness of the Cu film. The colour scale is from $-624$ pm (blue) to $+486$ pm (red), contour lines are dashed, solid lines are the lattice planes with distortions magnified. The residual standard deviations are 4 pm, 4 pm, and 6 pm, respectively.}\label{fig-05}
\end{figure}

The moir\'e topographies and displacement fields $u_{\rm M}$ are shown in Figs.\ \ref{fig-04} and \ref{fig-05}. The Cu films compress the crystal lattice and the stress increases with the thickness. This compression is consistent with the tensile stress of the Cu films (developing after growing and relaxation) reported in the literature \cite{Sharma:2015,Abadias:2018}.
\subsection{Surface-stress measurement}

To infer the magnitude of the tensile stress of the Cu film, we set up a finite element analysis of the interferometer mirror, modelled as a $(35\times 18\times 0.8)$ mm$^3$ Si crystal \cite{ElmerURL}. The effect of the film tensile-stress -- assumed equiaxial and uniform -- was simulated by a compressive surface-stress, $\tau(t_{\rm Cu})$, modelled as forces per unit length (from $\tau=1$ N/m to $\tau=4.5$ N/m, in variable steps), applied orthogonally to its 12 edges and lying in the crystal surfaces. Also, we set Dirichlet boundary conditions on the bottom surface, specifying null displacements, and used an anisotropic stiffness matrix \cite{Quagliotti_2013}.

\begin{figure}\centering
\includegraphics[width=0.9\columnwidth]{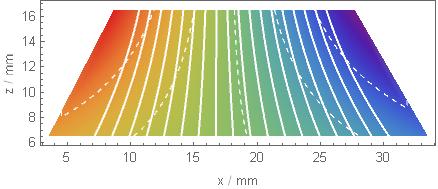}
\caption{Density plot (polynomial regression) of the simulated lattice displacements, $u_x(x,z;\tau=3.5\,{\rm N/m})$. The colour scale is from $-625$ pm (blue) to $+460$ pm (red), contour lines are dashed, solid lines are the lattice planes with distortions magnified. The abscissa and ordinate refer to the hypothetical mirror area imaged experimentally.}\label{fig-06}
\end{figure}

\begin{figure}\centering
\includegraphics[width=0.9\columnwidth]{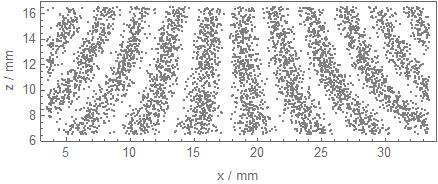}
\caption{Moir\'e topography inferred from the simulated displacements shown in Fig.\ \ref{fig-06}. The abscissa and ordinate refer to the hypothetical mirror area imaged experimentally. The fringe contrast has been set to one to improve visibility.}\label{fig-07}
\end{figure}

To realise a digital twin of the experimental set-up, the lattice displacement along the $x$ axis obtained via the finite element analysis, $u_x(x,z;\tau)$, was averaged over a pair of $(1 \times 3)$ mm$^2$ windows spaced by 4 mm, the separation of the x-ray paths at the interferometer mirror (see Fig.\ \ref{fig-01}). To simulate the experimental $(61 \times 8)$ images, the window pairs (the image of each scintillator pixel) were shifted vertically by eight 1.25 mm steps and horizontally by sixty-one 0.5 mm steps. Eventually, we found the polynomial regressions explaining the simulated data trading-off again between underfitting and overfitting according to \cite{Mana:2014,Mana:2019}. The regression of the displacement field for the $\tau=3.5$ N/m case is shown in Fig.\ \ref{fig-06}. Figure \ref{fig-07} shows the moir\'e topography inferred from this regression.

The similarities between the displacements and fringes depicted in Figs.\ \ref{fig-06} and \ref{fig-07} and the ones at the bottom of Figs.\ \ref{fig-04} and \ref{fig-05} are a clear hint that a surface stress of a few N/m is the quantity driving the observed strain of the crystal lattice. The strain magnitude is proportional to the density of the fringes, showing that the digital twin predicts correctly that the crystal is more strained at the free top and that the strain degrades at the bottom, where the crystal base hinders the lattice distortion. With a digital twin at hand that predicts the deformation of the crystal, we determined the magnitude and sign of the surface stress originated by the copper film and its dependence on the thickness.

\begin{figure}\centering
\includegraphics[width=0.95\columnwidth]{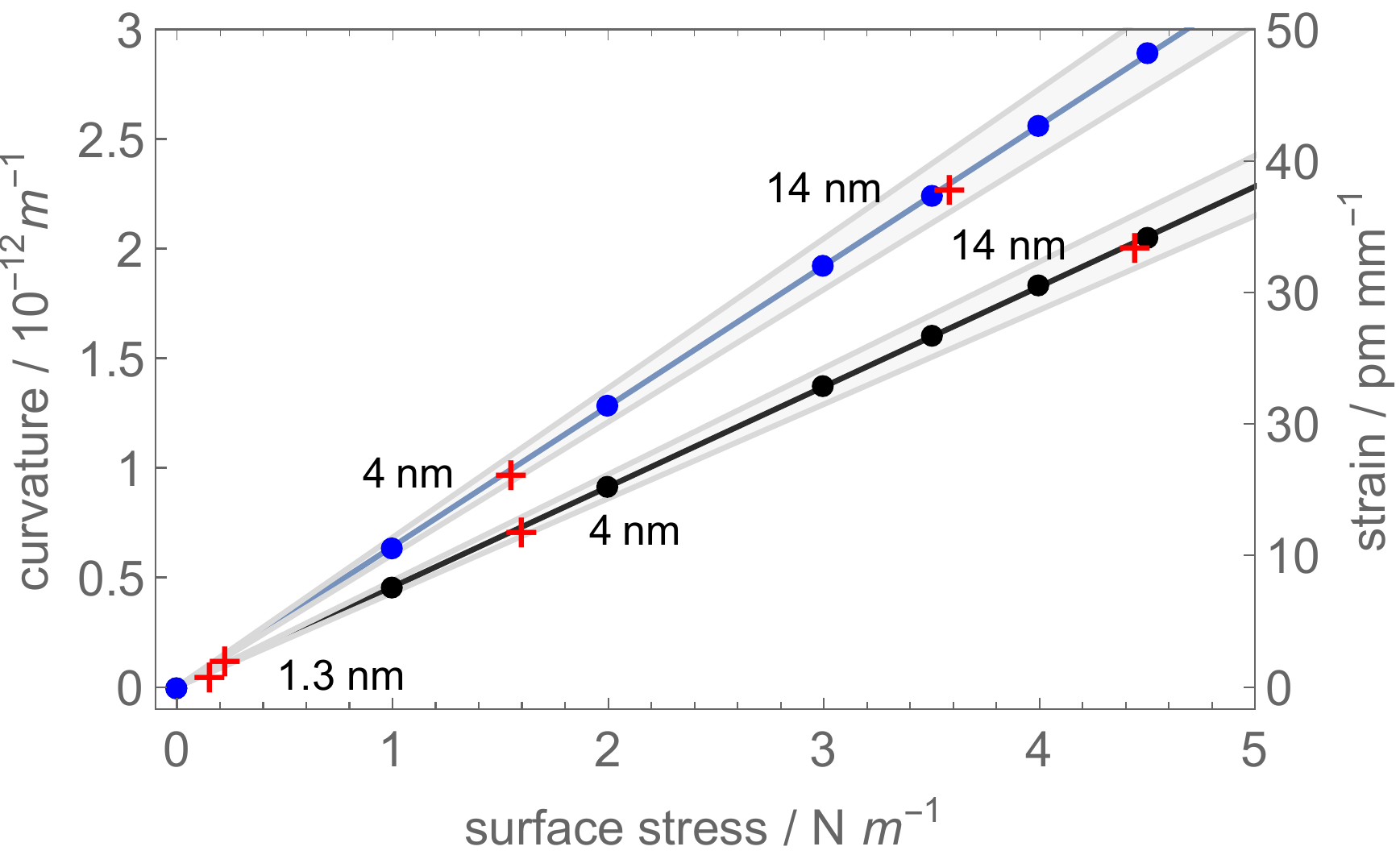}
\caption{Mean strains at the crystal top (black dots) and curvatures (blue dots) of the polynomial regressions and hyperbolic paraboloids best fitting the simulated lattice displacements. The red crosses are the intersections of mean strains and curvatures (obtained from the $t_{\rm Cu}=(1.3, 4, 14)$ nm experimental data) with the linear regressions best fitting the digital-twin data (solid lines). The filled areas represent the mean strains and curvatures when the mirror thickness varies from 0.75 mm to 0.85 mm.}\label{fig-08}
\end{figure}

The surface stress was estimated by using the mean strains observed at the mirror top -- $\bar{\epsilon}_{xx}^{\rm top}(t_{\rm Cu})=\Delta L/L$, where $L=30$ mm is the length of the moir\'e images -- in the nomogram shown in Fig.\ \ref{fig-08} (black line), which gives the same mean strains evaluated using the digital-twin data as a function of the surface stress. The values of the mean stress obtained via the optimal regression of the data,
\begin{equation}\begin{array}{lll}
                  \bar{\epsilon}_{xx}^{\rm top}(t_{\rm Cu}=1.3\; {\rm nm}) &=2.3(2)\;  &{\rm pm/mm} \\
                  \bar{\epsilon}_{xx}^{\rm top}(t_{\rm Cu}=4  \; {\rm nm}) &=16.0(2)\; &{\rm pm/mm} \\
                  \bar{\epsilon}_{xx}^{\rm top}(t_{\rm Cu}=14 \; {\rm nm}) &=37.0(3)\; &{\rm pm/mm}
                \end{array} ,
\end{equation}
where the standard uncertainties were estimated as $\sqrt{2}\sigma_u/L$ and $\sigma_u$ is the standard deviation of the residuals, led to
\begin{equation}\label{result-1}\begin{array}{lll}
                  \tau(t_{\rm Cu}=1.3\; {\rm nm}) &=0.22(2)\; &{\rm N/m} \\
                  \tau(t_{\rm Cu}=4  \; {\rm nm}) &=1.55(2)\; &{\rm N/m} \\
                  \tau(t_{\rm Cu}=14 \; {\rm nm}) &=3.58(3)\; &{\rm N/m}
                \end{array} .
\end{equation}

The simulated displacements are well approximated by (rectangular) hyperbolic paraboloids, having the axes rotated by 45$^\circ$ clockwise with respect to the $x$ and $z$ axes. They are quadratic surfaces given by the equation
\begin{equation}\label{eq:FittingPolyn}
 u_x(x,z;\tau) = u_0 - \kappa(\tau)(x-x_0)(z-z_0)/2 ,
\end{equation}
where $(x_0=14.1, z_0=-13)$ mm are the coordinates of the centre, $\tau$ is the surface stress, and $u_0$ and $-\kappa^2(\tau)$ are the displacement and Gaussian curvature in the centre.

The axes of the paraboloids approximating the experimental data might be slightly rotated to those approximating the simulated ones, and their centres displaced somewhat. While the $x$ and $z$ axes of the finite element model are parallel to the crystallographic directions (110) and (001), the axes experimentally determined might be rotated, might deviate from being orthogonal, and their origin might be displaced. Furthermore, the assumption of uniform surface stress might not be valid.

To accommodate these degrees of freedom, we observed that Gaussian curvature $-\kappa^2(\tau)$ is the only parameter of (\ref{eq:FittingPolyn}) depending on the surface stress. Also, as shown in Fig.\ \ref{fig-08}, $\kappa(\tau)$ depends linearly on $\tau$.

Furthermore, it is invariant under distance-preserving transformations allowing thus to compare misaligned paraboloids. This fact enables $\tau(t_{\rm Cu})$ to be estimated via the Gaussian curvatures of the paraboloids best fitting the experimental data.
Therefore, to take the possible misalignments between the set-up and finite-element model into account, we modelled the experimental images as
\begin{eqnarray}\label{model}\nonumber
 u_x(x,z) = a_{00} + a_{10}x + a_{01}z + a_{11}xz + a_{20}x^2 + a_{02}z^2 ,\\
\end{eqnarray}
and calculated the curvatures as $\sqrt{-|H|}$, where $H$ is the Hessian of (\ref{model}), and the vertical bars indicate the determinant. The results are
\begin{equation}\begin{array}{lll}
                  \kappa(t_{\rm Cu}=1.3\; {\rm nm}) &=0.07(3)\times 10^{-12}\; &{\rm 1/m} \\
                  \kappa(t_{\rm Cu}=4  \; {\rm nm}) &=0.73(3)\times 10^{-12}\; &{\rm 1/m} \\
                  \kappa(t_{\rm Cu}=14 \; {\rm nm}) &=2.03(6)\times 10^{-12}\; &{\rm 1/m}
                \end{array} ,\label{kappa}
\end{equation}
where the standard uncertainty of the (\ref{model}) residuals was propagated through the calculations also considering that, because of the pixel overlapping, the independent data are only a one-eight of the total. Eventually, we used the estimated curvatures in the $\kappa(\tau)$ nomogram, as shown in Fig.\ \ref{fig-08} (blue line), and found the abscissae
\begin{equation}\label{result-2}\begin{array}{lll}
                  \tau(t_{\rm Cu}=1.3\; {\rm nm}) &=0.16(7)\; &{\rm N/m} \\
                  \tau(t_{\rm Cu}=4  \; {\rm nm}) &=1.60(7)\; &{\rm N/m} \\
                  \tau(t_{\rm Cu}=14 \; {\rm nm}) &=4.44(13)\; &{\rm N/m}
                \end{array}
\end{equation}
corresponding to the curvatures listed in (\ref{kappa}). This procedure favours the $\tau(t_{\rm Cu})$ values that ensure the best overlapping (under a distance-preserving transformation) between the contour lines of the observed and simulated displacement fields.

The standard uncertainties associated with our estimates (\ref{result-1}) and (\ref{result-2}) take only the statistical dispersion of the phase data into account. A more realistic 10\% fractional uncertainty follows from the comparison of the estimates based on the mean strain and strain-field curvature.

\begin{figure}\centering
\includegraphics[width=0.9\columnwidth]{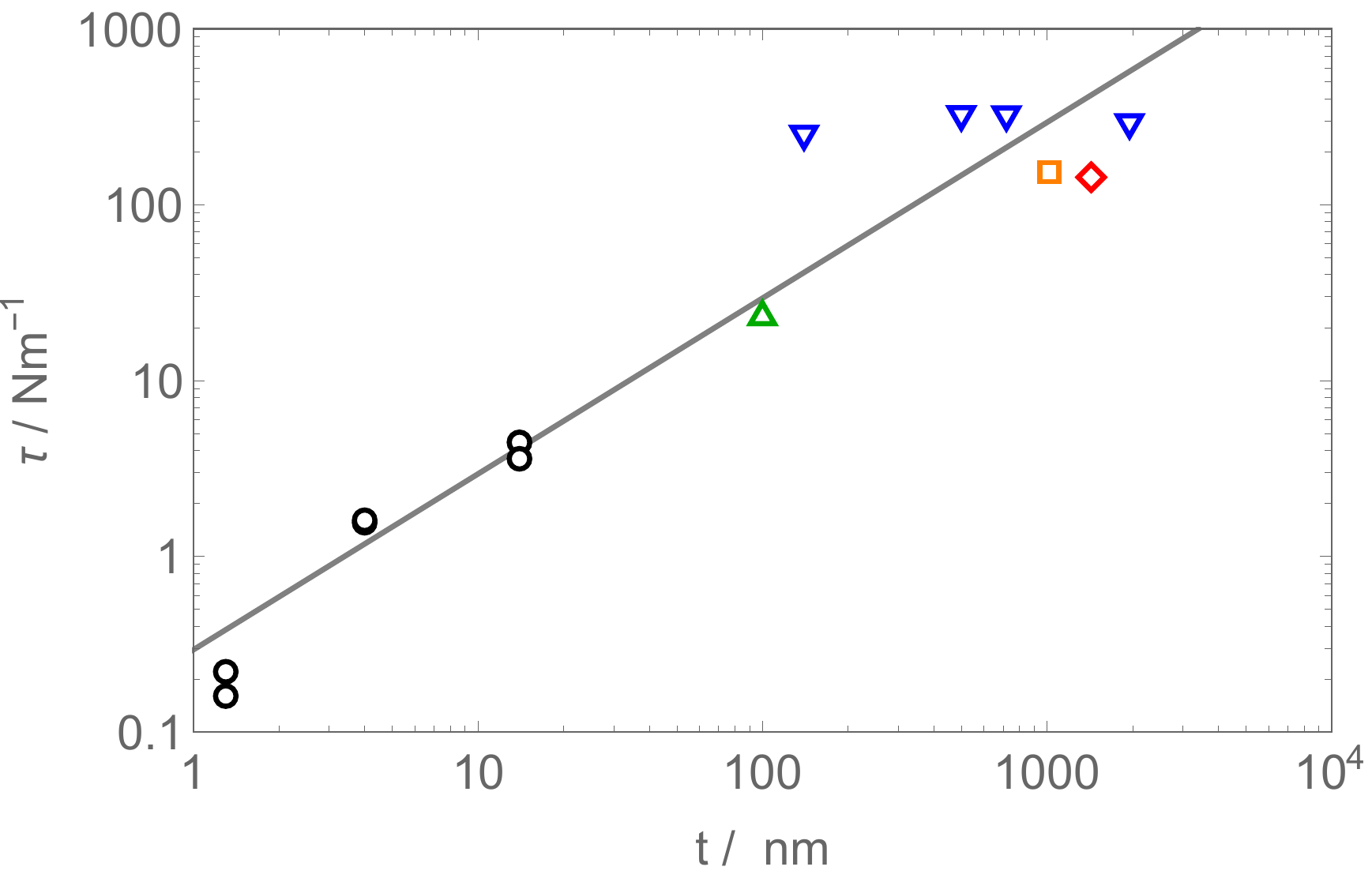}
\caption{Comparison of our surface stress values (black dots) with the stress-thickness of Cu films on acrylonitrile butadiene styrene (blue triangles), Ni-Fe alloy (orange square), Cu-Fe alloy (red diamond), and  SiO$_2$ (green triangle). Literature data are from \cite{Sharma:2015} (Fig.\ 4) and \cite{Abadias:2018} (Fig.\ 1). The black line is the linear regression best fitting our data.}\label{fig-09}
\end{figure}

A survey of the literature suggests that almost three orders of magnitude can be extrapolated from observed dependence of the surface stress on the film thickness. Figure \ref{fig-09} compares our surface stress values with the stress-thicknesses of Cu films on acrylonitrile butadiene styrene, metals, and SiO$_2$ \cite{Sharma:2015,Abadias:2018}. According to the linear regression of our data, the mean in-plane stress in the Cu film, 0.29(2) GPa, is independent of the film thickness.
\section{Conclusion}

The accurate measurement of the silicon lattice parameter by x-ray interferometry was a crucial step in counting the atoms in Si spheres -- via their unit-cell volumes -- for the determination of the Avogadro and Planck constants. It is now essential to the kilogram realisation, by reversing the count.

For a 1 kg Si sphere, the effect of the surface stress on the lattice parameter is irrelevant. However, it might be not so for the x-ray interferometer crystals, which, typically, are only 1 mm thick. This fact might harm the accuracy of the lattice parameter and unit-cell volume.

We have shown, via phase-contrast imaging of the crystal-lattice strains, that the surface finishing has measurable effects on the strain field of the diffracting planes, at the level of few parts per $10^9$.

We coated one of the interferometer crystals with nanometric Cu films, detected the strains of the diffracting planes, and determined the surface stresses explaining them. The resulting mean in-plane stress in the Cu films, 0.29(2) GPa, is consistent with the stress-thickness values given in the literature for similar interfaces \cite{Sharma:2015,Abadias:2018}. We also observe that the strains caused by nanometric Cu films support our fractional 1.25(72) nm/m correction of the measured lattice-parameter \cite{Bartl_2017,Fujii_2017}.

Phase-contrast imaging by x-ray interferometry can be used to determine the stress in different films on silicon. Nanometric SiO$_2$ films better represent the interface of the interferometer crystals, a reconstructed Si layer and an oxide film. Therefore, future work will aim at growing SiO$_2$ films having known thickness and at measuring the associated diffracting-plane strain.

\bibliographystyle{unsrt}
\bibliography{moire}
%\referencelist[moire]
\end{document}